\def\year{2022}\relax
\documentclass[letterpaper]{article} % DO NOT CHANGE THIS
\usepackage{aaai22 }  % DO NOT CHANGE THIS
\usepackage{times}  % DO NOT CHANGE THIS
\usepackage{helvet}  % DO NOT CHANGE THIS
\usepackage{courier}  % DO NOT CHANGE THIS
\usepackage[hyphens]{url}  % DO NOT CHANGE THIS
\usepackage{graphicx} % DO NOT CHANGE THIS
\usepackage{amsmath,amsfonts,amssymb, mathabx,bm,bbm}
\usepackage{xfrac}
\usepackage{tabularx, longtable}
\usepackage{natbib}  % DO NOT CHANGE THIS AND DO NOT ADD ANY OPTIONS TO IT
\usepackage{caption} % DO NOT CHANGE THIS AND DO NOT ADD ANY OPTIONS TO IT
\DeclareCaptionStyle{ruled}{labelfont=normalfont,labelsep=colon,strut=off} % DO NOT CHANGE THIS
\usepackage{algorithm}
\usepackage{algorithmic}

\usepackage{newfloat}
\usepackage{listings}
\floatstyle{ruled}
\newfloat{listing}{tb}{lst}{}
\floatname{listing}{Listing}

\usepackage{booktabs}

\usepackage{array}

\title{Siamese BERT-based Model for Web Search Relevance Ranking\\ Evaluated on a New Czech Dataset}
\author{
    Matěj Kocián\equalcontrib, Jakub Náplava\equalcontrib, Daniel Štancl\equalcontrib, Vladimír Kadlec
}
\affiliations{
    Seznam.cz, Prague, Czechia\\
    \url{{matej.kocian, jakub.naplava, daniel.stancl, vladimir.kadlec}@firma.seznam.cz}
}

\usepackage{bibentry}
\usepackage{eso-pic}% http://ctan.org/pkg/eso-pic

\begin{document}

\AddToShipoutPictureBG*{%
  \AtPageUpperLeft{%
    \hspace{0.5\paperwidth}%
    \raisebox{-\baselineskip}{%
      \makebox[0pt][c]{\textbf{This paper was accepted to IAAI 2022. Please reference it instead once published.}}
}}}

\maketitle

\begin{abstract}
Web search engines focus on serving highly relevant results within hundreds of milliseconds. Pre-trained language transformer models such as BERT are therefore hard to use in this scenario due to their high computational demands. We present our real-time approach to the document ranking problem leveraging a BERT-based siamese architecture. The model is already deployed in a commercial search engine and it improves production performance by more than 3\%. For further research and evaluation, we release DaReCzech, a unique data set of 1.6 million Czech user query-document pairs with manually assigned relevance levels. We also release Small-E-Czech, an Electra-small language model pre-trained on a large Czech corpus. We believe this data will support endeavours both of search relevance and multilingual-focused research communities.
\end{abstract}

\section{Introduction}

Web search engines are used by billions of people every day. Powered by results of decades of information retrieval research, they help find the documents people are looking for or directly answer their questions.

While basic query-document matching according to whether the documents contain all the words from the query might be sufficient for small document collections, the ever increasing quantity of documents available on the web makes it usually impossible for the user to go through all results that match given query words. Moreover, because of query-document vocabulary mismatch \cite{zhao_query_mismatch} and multiple possible word meanings, simple matching might exclude relevant documents. Therefore, there is a need for sophisticated natural language understanding (NLU) and document ranking methods. As the tasks might be intuitive for humans but difficult to describe algorithmically, such methods are usually based on machine learning utilizing examples provided by human annotators.

A popular document ranking model option is a Gradient Boosted Regression Trees (GBRT) %\cite{friedman_GBRT}
ranker \cite{zheng2007general}. It allows to easily and robustly combine hundreds of ranking features ranging from classical ones like BM25 \cite{robertson_BM25} or PageRank \cite{pagerank} to outputs of other statistical models. A number of features deal with the relevance of a document text to the query, which is basically a natural language processing (NLP) task.

Recently, the NLP community embraced BERT \cite{bert} inspired by the influential transformer architecture \cite{vaswani2017attention}. While BERT variants reach SoTA performance on many NLP tasks, they are computationally demanding and thus difficult to deploy in a search engine that strives to deliver results to users under a~second.

In this work, we create a new text relevance model based on Electra-small~\cite{electra} (a~variant of BERT) that improves relevance ranking while being sufficiently fast. We use the siamese architecture~\cite{sentence_bert} that allows us to precompute document embeddings and compare them with a query embedding at search time. We discuss several methods to compute the relevance score from the query and the document embeddings and propose a new neural-based interaction module. 

Most relevance research published so far deals with English queries and documents. We are interested in model performance on Czech data. To this end, we pretrain an Electra-small model on a Czech corpus and fine-tune it for relevance ranking on a Czech query-document dataset, which we also release to facilitate further research in this area.

Our main contributions are:
\begin{itemize}
    \item We develop and train an Electra-based siamese model for relevance ranking that has also been deployed in a search engine, where it improves performance by 3.8\%.

    \item We release DaReCzech\footnote{{\url{https://github.com/Seznam/DaReCzech}}}, a large Czech relevance dataset with real user queries and relevance annotations provided by human experts.
    
    \item We release Small-E-Czech\footnote{{\url{https://huggingface.co/Seznam/small-e-czech}}}, an Electra-small model pretrained on a Czech corpus. 
\end{itemize}

\section{Related Work}\label{sec:related_work}
This section provides an overview of related work. It describes transformer models, model compression and siamese transformers. The section is concluded with reviews of existing datasets for document ranking.

\subsection{Transformer models}\label{subsec:transformers}
Transformer model architecture, introduced by \citet{vaswani2017attention}, brought a revolution into NLP. They proposed an encoder-decoder model, intended for sequence transduction, based on a multi-head self-attention mechanism that enabled to learn long-term dependencies.

\citet{bert} introduced BERT, which was a novel encoder-only language model pre-trained on a large text corpus through masked tokens and next sentence prediction. Subsequently, the model was fine-tuned on a plethora of NLU tasks and reached SoTA results. Here, we rely on Electra~\cite{electra}, which shares its architecture with BERT, but it promises more efficient pre-training and it has been demonstrated that it can be trained in a smaller configuration than the one known as BERT-base (14M vs 110M parameters) without a dramatic drop in performance.

\subsection{Knowledge Distillation and Model Compression}\label{subsec:KD}
Knowledge distillation is a technique for transferring knowledge from large or ensemble models (teachers) to their smaller or single counterparts (students) \cite{bucila-model_compression}.
Current transformers, though SoTA, are prohibitively slow to  use in some settings, such as real-time web search. Therefore, many works have been dedicated to distilling knowledge to more compact models, e.g. \citet{distilbert} introduced DistilBERT, a compressed model with 6 layers, which resulted in $2.5\times$ speedup while retaining 97\% of the performance of BERT-base. 

During our work, we also distilled smaller variants of our Electra model having promising results. 
Although they provided us with a single-digit speed-up, calculating all query-document embeddings during online serving was still infeasible and we thus focus on siamese models.

\subsection{Siamese Transformers}\label{subsec:siamese}
Siamese architecture \cite{sentence_bert} is an orthogonal approach to speeding up online inference by offline pre-computation of document embeddings. In this setting, the model is separately fed two texts to obtain their embeddings. Subsequently, these two vectors are compared using e.g. cosine similarity to estimate a similarity score.

This approach was proved to be proficient in a first-stage document retrieval. \citet{repbert} computed the document relevance to a query as the scalar product of their embeddings and showed their BERT-based solution beat four traditional IR baselines. 

Similar approach with some additional adjustments was considered for ColBERT with likewise promising results \cite{colbert}. There, the similarity between a query and a document is evaluated over a bag of embeddings (i.e. there are multiple vectors for a query or a document). This, however, leads to high memory requirements as all embedding vectors need to be stored.

\citet{twinbert} presented TwinBERT, which is likely the closest work to ours. In that work, they first obtained query and document embeddings through [CLS] retrieved from the last BERT's layer. Afterwards, they compared the embeddings using an \textit{interaction module} which took an element-wise maximum of two embedding vectors and ran it through a residual fully-connected layer followed by a logistic regression layer to obtain the relevance score.

Our work differs from \citet{twinbert} in several aspects. (1) We use Electra instead of BERT due to more efficient pre-training. (2) We explore a deeper structure for the embedding interaction module. (3) We evaluate our model in the scenario of a web search instead of a sponsored search. (4) We fully focus on Czech, which is a much less resource-rich language than English. We release the manually annotated dataset to further support this research.

\begin{table*}[h!]
    \centering\footnotesize
    \begin{tabular}{lr|rrr|rrr|rrr|rrr|wr{0.8cm}wr{0.45cm}|wr{0.8cm}wr{0.45cm}}\toprule
     &  & \multicolumn{3}{c|}{Words per query} & \multicolumn{3}{c|}{Words per doc} & \multicolumn{3}{c|}{Words per title} & \multicolumn{3}{c|}{Docs per query} & \multicolumn{2}{c|}{Random} & \multicolumn{2}{c}{Oracle}\\
     Dataset & \#records  & \sfrac{1}{4} & avg & \sfrac{3}{4} &  \sfrac{1}{4} & avg & \sfrac{3}{4} & \sfrac{1}{4} & avg & \sfrac{3}{4} & \sfrac{1}{4} & avg & \sfrac{3}{4} & P@10 & DCG & P@10 & DCG\\
   
    \midrule
Train-big &  1\,431\,730 & 2 & 2.9 & 4 & 7 & 533.8 & 392 & 3 & 5.4 & 8 & 3 & 8.1 & 7 & 18.1 & 1.2 & 22.1 & 1.5\\
Train-small &  97\,386 & 2 & 3.0 & 4 & 1 & 300.3 & 198 & 2 & 4.5 & 6 & 37 & 52.6 & 65 & 36.2 & 6.9 & 82.2 & 8.2\\
Dev &  41\,220 & 2 & 2.9 & 4 & 2 & 310.7 & 218 & 2 & 4.5 & 6 & 36 & 52.0 & 66 & 34.9 & 6.7 & 80.4 & 8.0\\
Test &  64\,466 & 2 & 2.9 & 4 & 4 & 371.9 & 322 & 2 & 5.1 & 7 & 7 & 27.8 & 43 & 37.9 & 3.2 & 59.3 & 4.0\\
    \bottomrule
    \end{tabular}
    \caption{DaReCzech statistics. We report the number of words (whitespace separated) per extracted document body and title, number of annotated documents per query, and P@10 and Discounted Cumulative Gain (DCG) for random ranking (100 runs average) and ideal (oracle) ranking. For number of words and documents we report the mean and 0.25 and 0.75 quantiles.}
    \label{table:dataset_stats}
\end{table*}

\subsection{Review of Existing Datasets}\label{subsec:existing_dataset}
To the best of our knowledge, there is no annotated dataset in Czech for relevance ranking.
Also, many datasets for document retrieval tasks were collected several years ago and are therefore outdated. The non-exhaustive review of some of the most prominent datasets is provided below.

The dataset most related to ours is MS MARCO \cite{ms_marco}. This dataset contains a collection of 1\,M user queries, together with 8.8\,M passages retrieved from 3.6\,M web documents obtained by the Bing search engine. In contrast to ours, all data are in English. Another dataset based on the Bing search logs is ORCAS~\cite{orcas} containing 20\,M query-document pairs, although it lacks annotations for any relevance task.

TREC2009 Web Track \cite{trec2009overview} overviewed retrieval techniques, and was based on a large corpus of 10 billion web pages in 10 languages crawled in 2009 called ClueWeb2009.\footnote{ \url{http://lemurproject.org/clueweb09/}} TREC2009 consists of several tasks including ad-hoc search where the aim was to provide a list of most relevant documents for unseen topics.

Another two datasets (US and Asian versions) were published by Yahoo for a learning-to-rank challenge \cite{chapelle2011yahoo}. They consist of annotated query-document pairs accompanied with relevance labels. All queries originate from real Yahoo search logs.

\section{Problem and Data}\label{sec:problem_data}

\begin{figure}[!htb]
    \centering
    \includegraphics[scale=0.53]{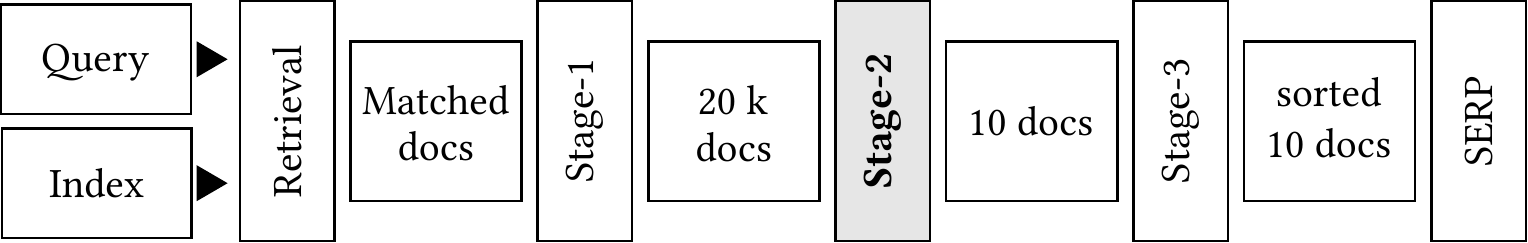}
    \caption{Ranking schema of the search engine. Indexed documents that match given query are evaluated by Stage-1 ranking model and top documents are sent to Stage-2, which we focus on. Stage-2 ranking model selects top 10 documents and sends them to Stage-3, which determines their final ordering on the search engine result page.}
    \label{fig:production_schema}
\end{figure}

For performance reasons, the document index currently has about 200 shards on 100 machines and the relevance ranking in the search engine consists of several stages (similar to those described by \citet{yahoopaper}, see Figure \ref{fig:production_schema} for an illustration). First, the retrieval stage selects documents containing all words from the original query or its enhanced variants (generated by typo correction, declension, etc.). Then the so-called Stage-1 selects about 20\,000 candidate documents using a GBRT ranker with fast features (PageRank, BM25 variants, etc.). In our research, we focus on Stage-2, which selects top 10 documents, again using a GBRT ranker. In addition to the features from Stage-1, Stage-2 uses also more complex ones (text relevance utilizing distances of query words matches in the document, etc.), totalling to more than 500 features. Finally, the top 10 documents are reordered by Stage-3. 

We improve Stage-2 by adding a new feature to the GBRT ranker. This is not easy as the ranking features have been tuned for years, and such efforts often result in negligible improvements.

The quality of the ranker is periodically evaluated on a set of about 2\,500 queries sampled from the past 3-month period of the query log. For each query, top 10 results are retrieved and their relevance is evaluated by human annotators. As the order of the top 10 results might be changed by Stage-3, we primarily measure Precision@10 (P@10), i.e.\ the ratio of relevant documents among the top 10.

After a new evaluation query set is sampled, the annotated query-document pairs from the last set are added to an old data pool and can be used for training and preliminary testing of models. Note there are much fewer annotated documents per query in the data pool than ca. 20\,000 candidates available in production Stage-2. Generally, these annotated documents must have been deemed relevant by a previously evaluated model. A substantially different model that would be able to bring new relevant documents to the top in production is thus at a disadvantage. We hence consider our test set only as an approximation of the final evaluation.

Another problem with old data is that documents might have changed (or their relation to the world, e.g.\ in case of current events, shifted word meanings, user expectations, etc.) and thus the relevance annotations might be outdated. This is the reason why the GBRT rankers are usually trained only on a recent subset of the old data pool. The rest can then be used for text features training, the rationale being that text content relevance might be less ephemeral.

\subsection{DaReCzech}
\begin{table*}[t!]
    \centering\footnotesize
    \begin{tabular}{lp{0.88\linewidth}}\toprule
    Field & Value\\
    
    \midrule
    %Query & musi byt test pri navratu z kanárské ostrovy\\[0.3cm]
    Query & volno otec po porodu\\[0.3cm]
    %URL & https://www.novinky.cz/cestovani/clanek/cestujici-na-kanarske-ostrovy-musi-od-14-listopadu-dolozit-negativni-test-40341157\\[0.3cm]
    URL & \url{https://www.seznamzpravy.cz/clanek/novinka-pro-cerstve-otce-tyden-placene-dovolene-po-narozeni-potomka-41487?autoplay=1}\\[0.3cm]
    Doc repr. & title: novinka pro čerstvé otce týden placené dovolené po narození potomka url: seznamzpravy.cz/clanek/novinka pro cerstve otce tyden placene dovolene po narozeni potomka 41487?autoplay=1 bte: Novinka pro čerstvé otce: týden placené dovolené po narození potomka Zapojení otců má pomoci matce v kritické fázi šestinedělí. A zároveň posílit vztah mezi dítětem a rodiči. Patří otcovská do ranku předvolebních dárků minulé vládní koalice? (\ldots)\\[0.3cm] %Nebo rodinám skutečně výrazně pomůže? A jak o ni požádat? Duel na Seznam Zprávy se ptal Pavla Krejčího, pověřeného ředitele České správy saociálního zabezpečení, a Jiřího Šatavy, analytika think tanku IDEA při CERGE EI. (\ldots)\\[0.3cm]
    Title & novinka pro čerstvé otce týden placené dovolené po narození potomka seznam zprávy\\[0.3cm]
    Label & 1.0\\
    \midrule
    \multicolumn{2}{c}{\textit{English translation}} \\ \midrule
    Query & \textit{father's leave after childbirth} \\[0.3cm]
    
    Doc repr. & \textit{title: news for fresh fathers a week of paid leave after the birth of offspring url: seznamzpravy.cz/clanek/news for fresh fathers a week of paid leave after the birth of offspring 41487?autoplay=1 bte: News for fresh fathers: a week of paid leave after the birth of the offspring The involvement of fathers is intended to help the mother at the critical stage of the six-week period. And at the same time strengthen the relationship between the child and the parents. Is paternity leave one of the last government coalition's pre-election gifts? (\ldots)} \\[0.3cm]  %Or will it actually help families significantly? And how to apply for it? Duel on Seznam Zprávy asked Pavel Krejčí, director in charge of the Czech Social Security Administration, and Jiří Šatava, analyst at the think tank IDEA at CERGE EI. (\ldots)} \\[0.3cm]
    
    Title & \textit{news for fresh fathers a week of paid leave after the birth of offspring} \\
    \bottomrule
    \end{tabular}
    \caption{Example dataset record with an English translation. The document representation was slightly shortened.}
    \label{table:dataset_example}
\end{table*}
DaReCzech is a new Czech dataset for text relevance ranking that we created from the old data pool.
% We created a text relevance dataset from the old data pool. 
It is divided into four parts: \textit{Train-big} comprising more than 1.4\,M query-document pairs (intended for training of a (neural) text relevance model used as a feature in the GBRT model),
%for language model\todo{JN: language model je obecne neco jineho; co treba training large (neural) models} fine-tuning),
\textit{Train-small} (97\,k records, intended for GBRT training), \textit{Dev} (41\,k records) and \textit{Test} (64\,k records), which contains the newest annotations.
There is no intersection between query-document pairs in the training, development and test data. The basic statistics of the dataset are presented in Table~\ref{table:dataset_stats}.

Each dataset record contains a query, a URL, a document title, a document representation and a relevance label.
The queries are real user queries with some typos corrected. A~document representation consists of three parts:
\begin{itemize}
    \item Title -- document title words that were classified by an internal model of the search engine as words corresponding to that particular document, as opposed to words corresponding to the whole web site (usually domain name or description). It is lowercased.
    \item URL -- a preprocessed document URL, with \% sequences decoded, plus signs converted to spaces and some parts (matching the regex \verb#(https?:\/\/(www\.)?|[-_\t])#) removed.
    \item Body Text Extract (BTE) -- document body filtered with an internal model of the search engine, i.e. supposedly without headers, menus, etc.
\end{itemize}
The processed parts are then prepended with identifiers and concatenated: \texttt{title: <title> url: <url> bte: <bte>}. 

The relevance labels were mapped from the original annotations as follows: \textit{(1) Useful}: 1,
\textit{(2) A little useful}: 0.5 (0.75 for \textit{Test}),
\textit{(3) Almost not useful}: 0.5 (0.25 for \textit{Test} and \textit{Train-big}),
\textit{(4) Not useful}: 0.
Note that because we track P@10, i.e. the ratio of useful (label $> 0.5$) documents among top 10, the exact values of other mapped annotations are less important on \textit{Dev} and \textit{Test} set.

For an example dataset record, see Table~\ref{table:dataset_example}. Some documents have empty bodies or titles, either because they did not contain any text in these fields or they were not indexed at the time of dumping the data from the search engine database. We dropped empty documents from the training set, as initial experiments showed  this helps the fine-tuning.

\subsection{Czech Corpus for Language Model Pretraining}
\label{ssec:czech_corpus_pretraining}
For self-supervised LM pretraining, we use an in-house Czech corpus (253~GB) that is once a year generated from documents downloaded by the search engine crawler. During the corpus generation, document language is detected, non-Czech, duplicate, SPAM and too short texts are dropped and the remainder is cleaned and lowercased.

\subsection{Baseline GBRT Ranker}
Relevance ranking in Stage-2 is done by a GBRT ranker using hundreds of features. The exact list changes over time as new features are implemented and old systems are turned off. In our work, we tried to improve a baseline model with 575 features. Examples of the most influential include:

\begin{itemize}
    \item dynamic text relevance -- scores depending on distances between matches of query words in the document, averaged across different generated query variants,
    \item PageRank,
    \item logistic regression using a query and title words as features,
    \item Okapi BM25 and its several variants.
\end{itemize}

\section{Model Architecture}\label{sec:model_architecture}

The core of our system is a Czech Electra model pretrained on the web corpus gathered by the search engine crawler. On top of this model, we build two alternative architectures: First, the \emph{query-doc model}, which uses a simple linear layer to transform the output of Electra's [CLS] token into a single number describing the relevance between the concatenated query and document. Second, the \textit{siamese model}, which uses the underlying Electra model to compute query and document embeddings. These embeddings are further compared using cosine similarity or a~small feed-forward network that outputs the~final relevance score.

The \textit{query-doc model} has a clear advantage over the \emph{siamese model} as it can directly compare subwords of both the document and the query. The \emph{siamese model}, on the other hand, has to encode all information about a query or a document in a vector of a limited size and compare these later. Nonetheless, at inference time, when the best document should be selected for a query, all query-document pairs need to be evaluated by the whole model for the \textit{query-doc} approach. This turned out to be computationally infeasible even in Stage-2 as 20\,000 document embeddings would have to be computed for each query.

In this section, we first describe the \textit{query-doc model} and the \textit{siamese model} architectures. We then elaborate on a set of improvements applied to the latter model to decrease the gap between the performance of these two systems while keeping the latency low. Finally, we describe the training details. 

\subsection{Query-Doc Model}

The \textit{query-doc model} follows the original approach for sequence classification  \cite{bert} by adding an additional linear layer on top of the Electra embedding for the artificial [CLS] token. We add a sigmoid activation to project scores between 0 and 1. The input to this  model is a single sequence: a tokenized query and a document representation separated by the special [SEP] token. The model outputs a number predicting the document relevance for the query.

\subsection{Siamese Model}

The \textit{siamese model} utilizes an underlying Electra model to compute embeddings separately for a query and a document. Similarly to \citet{sentence_bert}, we experimented with three strategies of whole token sequence embedding computation: mean or maximum of all output vectors or the output for the [CLS] token. We found the [CLS] token to work best. The embeddings are then compared using cosine distance serving as a relevance proxy.

\subsection{Improving the Siamese Model}

\begin{figure}[!htb]
    \centering
    \includegraphics[width=0.89\linewidth]{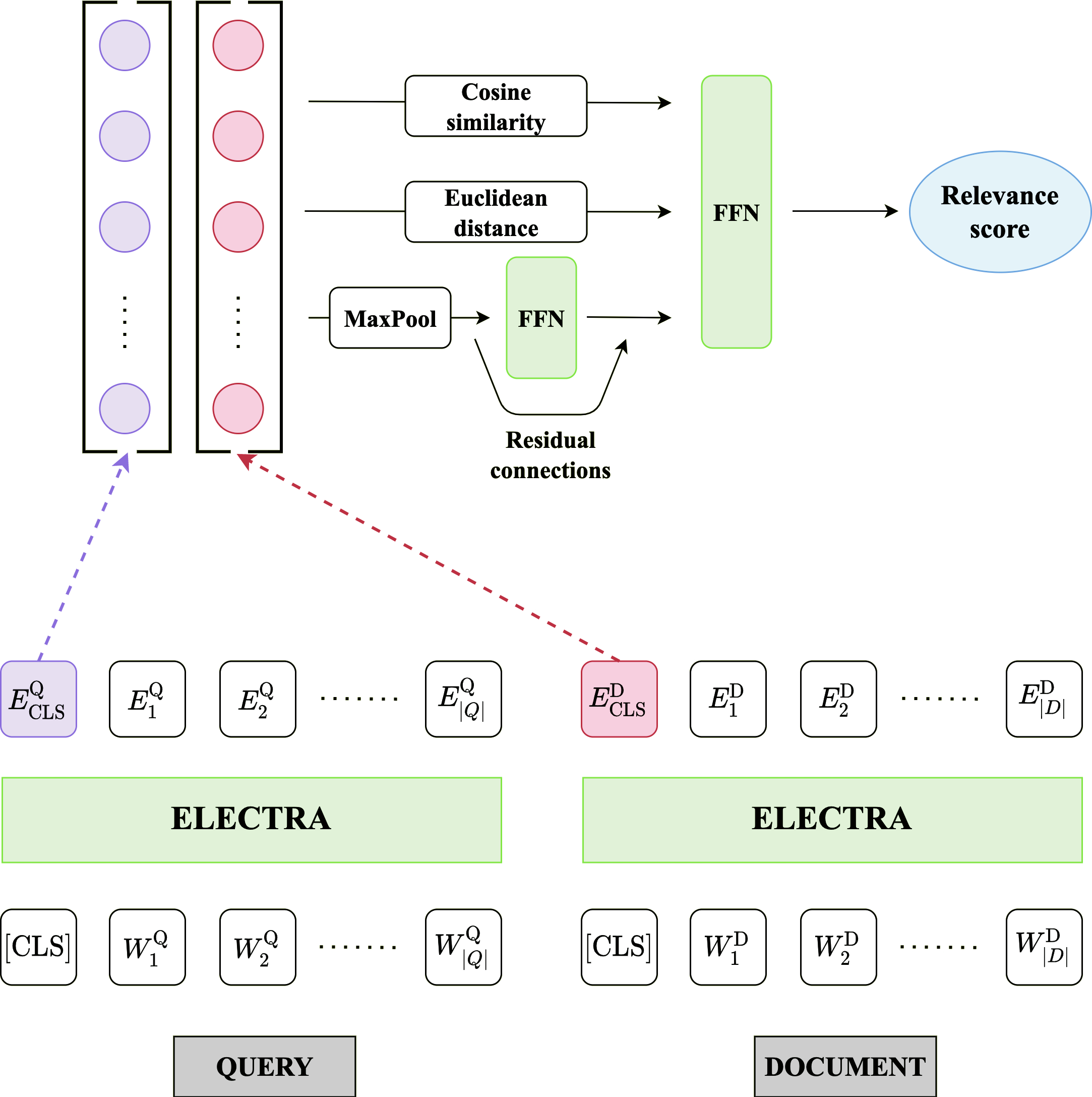}
    \caption{The final \textit{siamese model}. The tokenized query and document are inputted to Electra separately (tokens $W^Q_i$ and $W^D_i$), embeddings from their [CLS] tokens are compared using a custom interaction module. The module comprises a 2-layer feed-forward network and Euclidean distance and cosine similarity, followed by a linear transformation and hyperbolic-tangent non-linearity.}
    \label{fig:my_label}
\end{figure}

\subsubsection{Custom Interaction Module}
\label{ssec:custom_metric}

Cosine similarity has proven to be an effective and fast way to compare embeddings~\cite{sentence_bert}, but its simplicity might limit performance. Therefore, similarly to \citet{karpukhin2020dense}, we define a feed-forward network that compares the embeddings and returns a relevance score. The small size of the network ensures that it still remains fast enough.

Following \citet{twinbert}, the input to our interaction module is an embedding $e(q)$ of a query $q$ and an embedding $e(d)$ of a document $d$, each being of dimension $n = 256$. First, we compute the element-wise maximum of the input embeddings 
%$$ x = max([emb_q, emb_d], axis = 0)$$
$$m = \max(e(q), e(d)).$$
This is processed by two fully-connected layers inspired by the fully-connected block in the transformer model. The first layer maps the input vector to a space with twice as many dimensions and is followed by dropout (drop probability 0.25) and GELU activation \cite{gelu2016}. The second layer maps the vector back to the original space and again applies GELU. We also use a residual connection circumventing the nonlinearity:
\begin{align*}
h_1 &= \mathrm{Dropout_{0.25}}(\mathrm{GELU}(W_1 m)),\\
h_2 &= \mathrm{GELU}(W_2 h_1) + m,
\end{align*}
where $W_1 \in \mathbb{R}^{2n \times n}$ and $W_2 \in \mathbb{R}^{n \times 2n}$ are learnable weight matrices.
%$$ inv\_bottleneck\_hidden = Linear(256 -> 512, x) $$
%$$ inv\_bottleneck\_act = Dropout(Gelu(inv\_bottleneck\_hidden)) $$
%$$ inv\_bottleneck\_out = Linear(512 -> 256, inv\_bottleneck\_act)$$
%$$ inv\_bottleneck\_out\_act = GELu(inv\_bottleneck\_out) + x$$
The output $h_2$ of this residual block is concatenated with cosine similarity and Euclidean distance between the query and document embeddings. We found that this improves the stability of training.

$$h_3 = [h_2, \hspace{0.1cm} \cos(e(q), e(d)), \hspace{0.1cm} \Vert e(q) - e(d)\Vert]$$
%$$ features\_enhanced = [inv\_bottleneck\_out\_act, cos\_sim(emb_q, emb_d), euc_sim(emb_q, emb_d)]$$
Finally, a linear layer with a \emph{tanh} activation is used to produce the final relevance score: %(we use hyperbolic tangent to match the output range of cosine similarity):
$$r = \tanh(w_\mathrm{out} \cdot h_3),$$
where $\ w_\mathrm{out} \in \mathbb{R}^{n + 2}$ is a learnable weight vector.
%$$ relevance = \tanh(\mathrm{Linear}(256 \to 1, features\_enhanced)$$

\subsubsection{Considering Multiple Electra Layers}

\citet{tenney-etal-2019-bert} have shown that different tasks benefit more from different layers of BERT. Following the approach of \citet{kondratyuk201975}, we do not use only the last-layer embedding of the [CLS] token, but learn a weighted combination of all layer outputs for this token and take that as the embedding of the input sequence.

\subsubsection{Learning with a Teacher}
The query-doc model performs better than the siamese one, but is impractical to deploy due to its computational demands. Therefore, we use a variant of knowledge distillation to bridge this gap in quality.

Specifically, for each training sample, we compute the prediction of the query-doc (teacher) model, average it with the original label and use the result as a training label for the siamese (student) model.

\subsubsection{Initialization from the Teacher.} We initialize the student model weights with the fine-tuned
teacher weights.

\subsubsection{Ensemble}

Ensembling multiple models (i.e. combining their outputs) proves to improve results at the cost of increased inference time \cite{ensemble_ml}. Having a fast enough siamese model, we found out that having two models in an ensemble is a viable option. To diversify the models, only the random seed was changed when training the second one. This affected the initialization of the interaction module weights, the order of training samples and dropout.

We tried combining outputs of the models by taking either the mean or the maximum prediction and found the former to work better.

\subsection{Pretraining Small-E-Czech}
An internal 253~GB Czech web corpus was used for unsupervised pretraining. The texts are tokenized into subwords with a standard BERT WordPiece tokenizer~\cite{schuster2012japanese}. The tokenizer is trained on a subset of the corpus and its vocabulary size is limited to 30\,522 items. 

The Electra model is pre-trained using the official code\footnote{\url{https://github.com/google-research/electra}} in the Electra-small configuration. We train the model for 4\,M training steps, which took ca. 20 days on a single GPU.

\subsection{Training Details}

We train our model on the \textit{Train-big} set and select the best checkpoint using early-stopping on the \textit{Dev} set. Subsequently, we train a GBRT ranker on the \textit{Train-small} set with our model output as an additional feature and evaluate both on the \textit{Test} set. All input texts are lowercased to match the pretraining corpus.

We use Adam optimizer with learning rate $5\cdot10^{-5}$ without any warmup or learning rate decay to optimize weights of the Electra model and a custom interaction module if present. We use MSE loss for the query-doc and the siamese models. We also experimented with other loss functions such as \textit{triplet} loss, but found them to perform worse.

We cap each sentence at 128 tokens and train with batches of size 256. For siamese models, we map the labels into $[-1, 1]$ to match the model output range.

For knowledge distillation, our loss function is a mean of MSE between student and teacher prediction (soft labels) and conventional MSE with respect to (hard) gold labels. Otherwise, all training parameters remain the same.

We code our experiments using PyTorch \cite{pytorch2019} and the Transformers library \cite{transformers2020}.

The GBRT ranker is trained using the Catboost library with 1\,500 trees of depth 6, RMSE loss function and early stopping on 100 iterations.

\section{Results}\label{sec:results}

In this section, we present the results of training our \emph{query-doc} and \emph{siamese} models. We train each model 4 times with different random seeds (affecting the initialization of the custom interaction module if present and the dropouts), select the best checkpoint for each run on the development set and report the mean and the standard deviation of the 4 runs on the test set.

We report two types of results – the first one labeled as \textit{Standalone} for the respective model being used alone for ranking; and the second one labeled as \textit{with GBRT} for the respective model being used as an additional feature for a GBRT ranker that already utilizes hundreds of existing features. Note that we use the \textit{Train-big} data to train the neural models and, subsequently, the \textit{Train-small} data to train the GBRT ranker with the exception of the production search engine baseline that is trained on the entire \textit{Train-big} data.

We evaluate the models in two scenarios: (1) on the new DaReCzech dataset, (2) in a production setting.

\subsection{DaReCzech}

Table~\ref{table:results_main} presents an evaluation on DaReCzech dataset. In~the top part, we show results of baselines – the random ranking, BM25 and the production GBRT ranker (\textit{Search engine baseline}), and P@10 achievable by ideal ranking (\textit{Oracle}).

The \textit{query-doc model} outperforms the baseline results by a large margin, achieving P@10 46.3 and GBRT P@10 46.93. 
These results set the upper bound for the siamese model as the query-doc approach may compare tokens of both query and document directly.

\begin{table}[!htb]
    \centering\footnotesize
    \begin{tabular}{lll}\toprule
    & \multicolumn{2}{c}{Precision@10}\\
    Model & Standalone & with GBRT \\\midrule
    Random Baseline & 37.90 & -- \\
    BM25 & 40.47 & -- \\
    Search engine baseline & -- & 45.14 \\
    Oracle & 59.30 & -- \\\midrule
    
    Query-Doc & 46.30 $\pm$ 0.17 & 46.93 $\pm$ 0.12 \\\midrule
    
    Siamese-Cosine & 42.46 $\pm$ 0.15	& 45.41 $\pm$ 0.14 \\
     + custom inter. mod. & 43.82 $\pm$ 0.45 & 45.90 $\pm$ 0.17 \\
     + weighted CLS & 44.72 $\pm$ 0.39	& 46.02 $\pm$ 0.18 \\
     + knowledge distillation & 45.00 $\pm$ 0.36 & 46.26 $\pm$ 0.19\\
     + teacher initialization & 45.26 $\pm$ 0.22 & 46.42 $\pm$ 0.14\\
     + ensemble (2 best) & 45.49 & 46.61  \\\bottomrule
    \end{tabular}
    \caption{Results on DaReCzech. For each model / additive improvement, we report Precision@10 of the model and the GBRT ranker with the model output as an additional feature.}
    \label{table:results_main}
\end{table}

Despite its relative simplicity, the feature from the \textit{Siamese-Cosine} model helps the GBRT ranker by ca. 0.3 percent, but is not very competitive when used alone, and even with the GBRT ranker it lags behind the query-doc approach. When the cosine distance is replaced with a more sophisticated neural based interaction module, the performance improves, and this modification appears as the strongest one.

Using a weighted combination of multiple Electra layers instead of the last layer output seems to improve the performance. However, we found that this may be due to our choice of the interaction module. When the weighting is used with the simplest model with the cosine similarity, it increases its performance only by ca. 0.2.

Both knowledge distillation from a query-doc teacher and weight initialization from the teacher help the model.

All described improvements to the baseline model proved to be effective. Their combination and the final ensembling reduced the gap between the siamese and the query-doc model greatly. Moreover, we can see that already our best non-ensemble siamese model has better performance (45.26) than the baseline production GBRT ranker (45.14). When we add the ensemble output to the features and retrain the GBRT ranker, its P@10 increases by 1.48 to 46.61.

\subsection{Real Traffic}
Model evaluation on a fixed test set is cheap and stable, but does not take into account the multitude of documents retrieved for a query in production from which the model can select the top 10. To account for this, 3\,000 queries were sampled from the search log. Top 10 documents were retrieved for each using the original GBRT ranker and the new GBRT ranker utilizing new Electra ensemble signals as additional features. The query-documents pairs were then assigned relevance levels by human experts. The new features increased P@10 of the model by 3.8\% (relative).

\section{Ablation Studies}\label{sec:ablation_studies}

In this section, we present several ablation studies. First, we inspect the importance of individual document parts. We then explore the effect of training data volume on model performance. Third, we study different interaction modules. Fourth, we evaluate a different initialization of the underlying Electra model and also experiment with bigger underlying models. Finally, we present model quantization results.

\subsection{Document Representation}

The document is represented using its title, URL and BTE. To explore the individual effects of these parts on model performance, we trained a different \textit{siamese} model on each part. No teacher was used during the training, because this would require training the teacher on the respective data part as well, i.e. we used \textit{+weighted CLS} model configuration from Table~\ref{table:results_main}. The testing was then performed on the test set comprising only the respective data part. The results of this experiment are displayed in Table~\ref{table:dataset_single_parts}. We can see that BTE contains the most useful information, but all data parts are useful, as the respective models are significantly better than the random baseline of 37.9 P@10 (see Table~\ref{table:dataset_stats}). 

Moreover, we conducted an experiment where the individual data parts are added incrementally, i.e. title, URL and BTE. The results are shown in Table~\ref{table:dataset_additive_parts}.

\begin{table}[!htb]
    \centering\footnotesize
    \begin{tabular}{lcc}\toprule
    & \multicolumn{2}{c}{Precision@10}\\
    Model & Standalone & with GBRT \\\midrule
    Title & 42.73 $\pm$ 0.09 & 45.46 $\pm$ 0.08\\
     %URL & 41.395 $\pm$ 0.63 & 45.37 $\pm$ 0.15 \\
     URL & 41.40 $\pm$ 0.63 & 45.37 $\pm$ 0.15 \\
     %BTE & 43.745 $\pm$ 0.46 & 45.76 $\pm$ 0.10 \\
     BTE & 43.75 $\pm$ 0.46 & 45.76 $\pm$ 0.10 \\\bottomrule
     % BWLinks & 42.98 $\pm$ 0.51 & 45.40 $\pm$ 0.01  \\\bottomrule
     %BWLinks & 42.98 $\pm$ 0.51 & 45.395 $\pm$ 0.01  \\\bottomrule
    \end{tabular}
    \caption{Effect of using only a single data part (no teacher).}
    \label{table:dataset_single_parts}
    
    \vspace{0.5cm}
    \begin{tabular}{lcc}\toprule
    & \multicolumn{2}{c}{Precision@10}\\
    Model & Standalone & with GBRT \\\midrule
    Title & 42.73 $\pm$ 0.09 & 45.46 $\pm$ 0.08\\
     + URL & 43.74 $\pm$ 0.37 & 45.84  $\pm$ 0.17 \\
     + BTE & 44.72 $\pm$ 0.39  & 46.02 $\pm$ 0.18 \\\bottomrule
     % + BWLinks & 44.73$\pm$ 0.32 & 46.13 $\pm$ 0.15 \\\bottomrule
    \end{tabular}
    \caption{Effect of using different subsets of document parts (cumulative, no teacher).}
    \label{table:dataset_additive_parts}

\end{table}

\subsection{Training Data Volume}

\begin{figure}[!htb]
    \centering
    \includegraphics[width=0.9\linewidth]{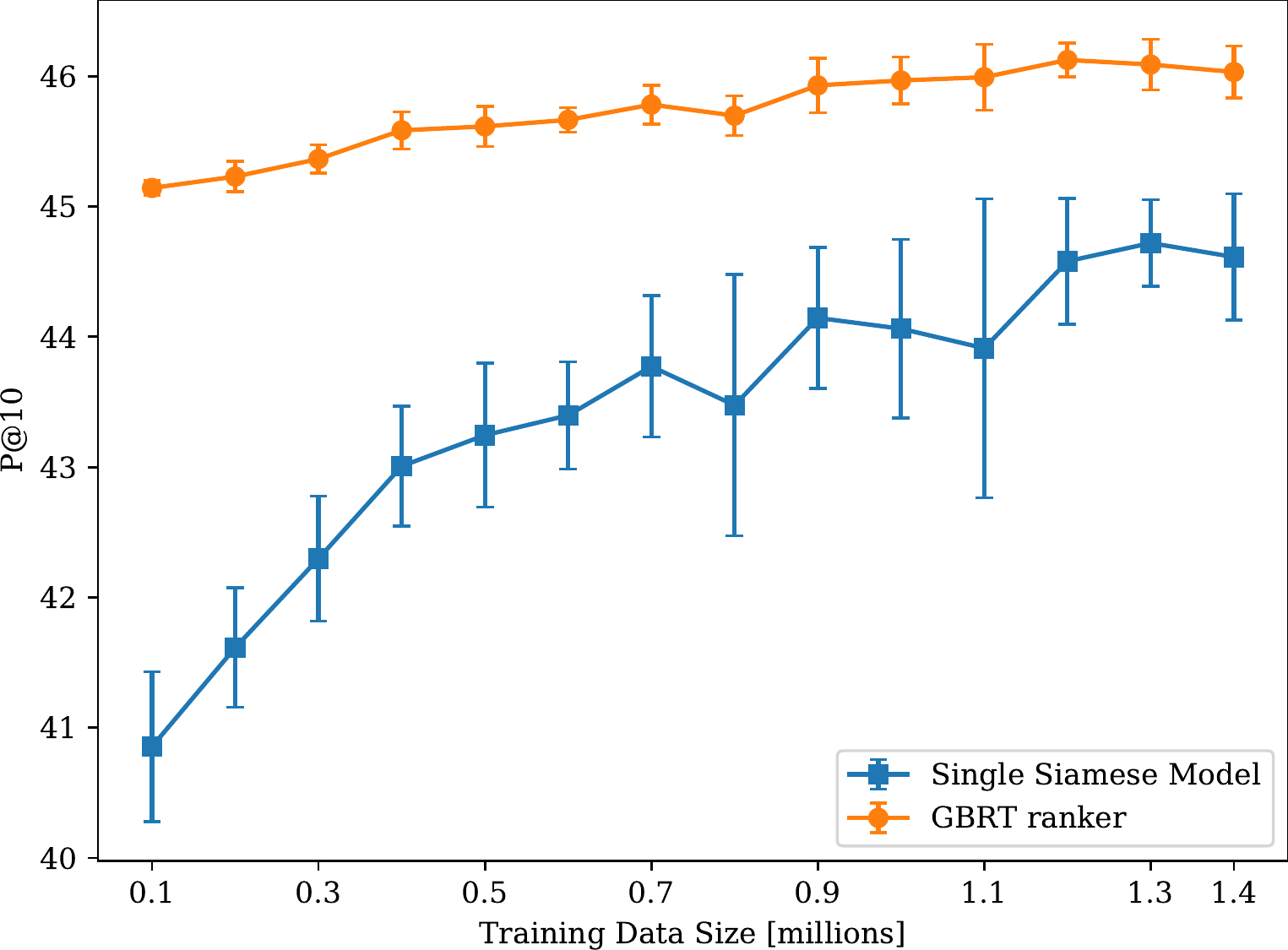}    
    \caption{Precision@10 of the model when trained only on a subset of the training data of particular size. We report the performance of the sole model and also of the GBRT ranker using the model output as an additional feature.} %We run each experiment 4 times on different subsets and report the standard deviation of the runs.}
    \label{figure_train_size}
\end{figure}

We inspect the effect of the number of training samples on model performance in Figure~\ref{figure_train_size}. Specifically, for each predefined training set size, we randomly sample this amount of data from the training set and train a \textit{siamese model} on it. We do not use a teacher and run each experiment four times to account for the randomness in sampling. The results show that the performance increases with the number of training samples, both of the model alone and the GBRT ranker using model output as an additional feature, while the gap between them decreases. The effect on performance slowly levels off, but the model might still benefit from more data.

\subsection{Interaction Module Variants}

As we already discussed in Section \textit{Custom Interaction Module}, the interaction module comparing two embeddings and returning a single relevance score may be cosine similarity or a feed-forward neural network. The final interaction module we use is a result of several preliminary experiments. We compare here five different architectures: 
\begin{itemize}
    \item Cosine -- compares the query and document embedding using cosine similarity
    \item Single Hidden -- a neural network mapping the query and document embeddings into a vector of size 3, concatenating it with their Euclidean distance and cosine similarity and finally using a simple feed forward layer with sigmoid activation to obtain the relevance score
    \item TwinBERT interaction module as proposed by \citet{twinbert} and described in Section \textit{Siamese Transformers}. Additionally, we use a weighted combination of token embeddings from different layers as it turned out to consistently improve performance.
    \item Final w/o cos/Euc -- our final interaction module as described in Section \textit{Custom Interaction Module} but without cosine similarity and Euclidean distance concatenated to the last hidden layer.
    \item Final -- our final interaction module as described in Section \textit{Custom Interaction Module}
\end{itemize}

\begin{table}[!htb]
    \centering\footnotesize
    \begin{tabular}{lccwc{1.0cm}}\toprule
     & \multicolumn{2}{c}{Precision@10} & \\
    Model  & Standalone & with GBRT & Speed-up \\\midrule
    Cosine & 42.46 $\pm$ 0.15	& 45.41 $\pm$ 0.14 & $2.7\times$ \\
    Single Hidden & 44.37 $\pm$ 0.17  & 46.06 $\pm$ 0.08 & $1.8\times$\\
    TwinBERT &  45.09 $\pm$ 0.17 &	46.22 $\pm$ 0.11 & $1.5\times$ \\
    Final w/o cos/Euc & 45.09 $\pm$ 0.16 & 46.30 $\pm$ 0.09 & $1.4\times$ \\
    Final &  45.26 $\pm$ 0.22 & 46.42 $\pm$ 0.14 & $1.0\times$ \\\bottomrule
    \end{tabular}
    \caption{Performance of the systems utilizing different interaction modules. Speed-up measurements regard the sole siamese model, not the GBRT.}
    \label{table:ablation_custom_metrics}
\end{table}

Table~\ref{table:ablation_custom_metrics} presents results and also relative speed-ups of the considered interaction modules. We can see that the better the model quality, the worse the model speed. 
The simplest cosine similarity is the fastest way to compare embeddings, but it performs the worst.
On the other hand, our final interaction module surpasses the performance of all other approaches, but is the slowest one. Still, depending on the document length, using the custom metric on top of the precomputed embeddings is roughly 1000$\times$ faster than running the entire query-doc model.

Two other noteworthy points are that using the Euclidean and cosine distances as additional features provides a slight gain in the final score, and that our final model surpasses the original TwinBERT interaction module.

% This indicates the need for a strong/expressive interaction module between the query and document embeddings.

\subsection{Base Models}

We decided to use Electra-small model due to its small size and high performance. Apart from the Electra-small model pretrained on Czech web documents, we experimented with three other base models:
\begin{itemize} 
    \item Electra-small model with the same vocabulary but initialized randomly 
    \item mBERT~\cite{bert} -- a well-known multilingual BERT language representation model
    \item RobeCzech~\cite{straka2021robeczech} -- Roberta-base model trained on Czech texts 
\end{itemize}

\begin{table}[!htb]
    \setlength{\tabcolsep}{0.27em}

    \centering\footnotesize
    \begin{tabular}{lc|wc{1.05cm}wc{1.15cm}|wc{1.05cm}wc{1.15cm}}\toprule
    & & \multicolumn{4}{c}{Precision@10} \\
     &  & \multicolumn{2}{c|}{Query-Doc} & \multicolumn{2}{c}{Siamese} \\
    Model & Params. & Standal. & w. GBRT & Standal. & w. GBRT \\\midrule
    Electra (rand.) & 13\,M & 44.21 & 45.67 & 41.55 & 45.39 \\
    Electra & 13\,M & 46.30 & 46.93 & 42.46 & 45.41 \\
    mBERT & 167\,M &  46.07 & 46.70 & -- & -- \\
    RobeCzech & 125\,M & 46.73 & 47.25 & 40.01 & 45.20  \\\bottomrule
    \end{tabular}
    \caption{Precision@10 of using different underlying BERT-based models. We report both results when trained in the query-doc and in the siamese mode. For simplicity, siamese models are trained with cosine similarity and without a teacher.}
    \label{table:other_base_models}
\end{table}

We trained all models in the \textit{query-doc} setting. As can be seen in Table~\ref{table:other_base_models}, the RobeCzech model performs the best, but is ca. $10\times$ bigger than our Electra-small model. We can also see that despite the relatively large finetuning dataset, the pretraining on monolingual data is still beneficial as the pretrained model outperforms the not-pretrained model.

In the \textit{siamese} mode, we trained all models except for mBERT which we omitted as RobeCzech provided better results in the query-doc setting. We use only cosine similarity as the embedding interaction module. Although the results show a big performance gap between Electra-small models and RobeCzech model, we think that the RobeCzech model would require more tuning of the learning rate schedule and other hyperparameters to fully exploit its capabilities.

\subsection{ONNX and Quantization}

Apart from using siamese architecture and an Electra-small variant, we tried to speed up our model using ONNX runtime\footnote{\url{https://github.com/microsoft/onnxruntime}}
and model quantization \cite{polino2018quantization}, i.e. reducing the precision of the computation. While our approach allows to precompute document embeddings offline, there are billions of documents in the database and generating embeddings can thus take a lot of time. We measured different combinations of ONNX conversion and quantization of the embedding module or the interaction module in Python using one thread on a CPU with one AVX-512 FMA unit. The results are in Table \ref{tab:quantization}. The interaction module running on ONNX with UINT8 model quantization is about $1.9\times$ faster than the Pytorch version, while the difference in quality is small. As for the embedding model, the difference in both speed and quality is bigger.

\begin{table}[!h]
    \centering\footnotesize
    \begin{tabular}{llcc}
    \toprule
        Embedding model & Interaction module & P@10 & Speed-up \\\midrule
        Pytorch FP32 & Pytorch FP32 & 45.27 & $1.0\times$\\
        Pytorch FP32 & ONNX FP32 & 45.27 & $1.2\times$\\
        Pytorch FP32 & ONNX UINT8 & 45.26 & $1.9\times$\\\midrule
        Pytorch FP32 & Pytorch FP32 & 45.27 & $1.0\times$\\
        ONNX FP32 & Pytorch FP32 & 45.27 & $1.5\times$ \\
        ONNX UINT8 & Pytorch FP32 & 45.13 & $3.0 \times$\\\bottomrule
    \end{tabular}
    \caption{Effect of model quantization on quality and speed. Relative speed-up values shown in the top part refer to the interaction module execution time and values in the bottom part refer to the embedding model execution times.}
    \label{tab:quantization}
\end{table}

\section{Model Size Effect on Response Times}

The query evaluation time depends on many factors, making it complicated to evaluate meaningfully. To give rough estimates, the query preprocessing phase gets prolonged by 10 ms on average when using the new Electra-small model. If we replaced it with a BERT-base model, the query embedding generation would take ca. 64 ms instead of 10 ms.

The retrieval and ranking phase used to take about 133 ms. With our new feature included, the computation takes about 136 ms (+3 ms) on average. Replacing Electra-small embeddings of size 256 with BERT-base embeddings of size 768 is expected to slow down the ranking stage to 143 ms (+10 ms).

\section{Conclusion}\label{sec:conclusion}
In this work, we presented a strong and fast variant of a siamese model for relevance ranking based on an Electra language model. We described and evaluated a set of improvements to the baseline siamese model and showed their effect on overall model performance. The model was successfully deployed as an additional feature for a GBRT ranker in a commercial search engine and led to a substantial improvement of 3.8\% in quality.

Moreover, we released Small-E-Czech, a pretrained Electra-small model, and DaReCzech, a new dataset for text relevance ranking in Czech. The dataset consists of more than 1.6\,M annotated query-documents pairs, which makes it one of the largest available datasets for this task.

\section{Acknowledgements}

We thank the developers and product managers who helped to put our prototype into production. Namely, Jaroslav Gratz, Aleš Kučík, Daniel Mészáros, Martina Pomikálková, Jakub Šmíd and Petr Vondrášek. We also thank the annotators who annotated the DaReCzech dataset and Ondřej Dušek and the anonymous reviewers for their valuable comments.

\footnotesize\bibliography{kocian}

\end{document}